# Low-frequency vortex dynamic susceptibility and relaxation in mesoscopic ferromagnetic dots

K. Yu. Guslienko[*]

*Material Science Division, Argonne National Laboratory, 9700 S. Cass Ave., Argonne, Illinois 60439*

**Abstract**

Vortex dynamics in a restricted geometry is considered for a magnetic system consisting of ferromagnetic cylindrical dots. To describe the vortex dynamic susceptibility and relaxation the equation of motion for the vortex center position is applied. The dependencies of the vortex dynamic susceptibility and resonance linewidth on geometrical parameters are calculated. A new method of extracting damping parameter from the vortex low-frequency resonance peaks is proposed and applied for interpretation of resonance data on FeNi circular dots.

---

[*] E-mail address: gusliyenko@anl.gov

Recently patterned magnetic mesoscopic and nano-structures have received considerable attention due to their possible applications in spintronics, magnetic field sensing and high-density information storage.[1] These structures, depending on their sizes and geometry, reveal either a single-domain or a vortex magnetization state in remanence. The questions of how fast magnetization switching occurs and for how long a given magnetization state can be stable with respect to thermal fluctuations are the central questions for different applications. The corresponding characteristic times (switching and relaxation times) are determined by a set of physical parameters, where the damping constant is a key parameter.[1,2] It is well established that a vortex state appears for large enough in-plane particle (dot) sizes.[3-5] Magnetic vortices being a kind of solitons[6] can be considered as topological excitations in ferromagnetic ordered media. The vortex is a strongly inhomogeneous state of the 3D magnetization field **M(r)** that cannot be reduced to the uniform **M** by any finite deformation. The vortex consists of a core with magnetization that is perpendicular to the dot plane and a main part with an in-plane flux-closure magnetization distribution.[4] There is a complex spin excitation spectrum over the vortex ground state. In particular, low-frequency translation modes of displacement of the vortex as a whole appear.[7,8] The magnetostatic interaction plays an important role in consideration of the vortex state stability and dynamic excitations in finite magnetic dots.

The vortex low-frequency excitation modes were observed by time-resolved Kerr,[9,10] X-ray[11-13] and strip line techniques.[14] Despite numerous experimental observations of magnetic vortices, there are few calculations of the dynamics of vortices in mesoscopic dots. The theory of vortex excitations in the dots was developed in Ref. 8, 15, 16, and numerical simulations were conducted in Ref. 15, 17, 18. The vortex translation, and high-frequency (radial and azimuthal) modes have been explored. However, the vortex state dynamic susceptibility and relaxation have yet not been considered for cases involving a restricted geometry.

In this Letter investigations of the vortex response to variable magnetic field stimuli are presented using as an example cylindrical magnetic dots. We consider dots with radius $R$, thickness $L$, and saturation magnetization $M_s$ (Fig. 1). The dot thickness is assumed to be about the same as the exchange length of the material, which allows one to neglect the dependence of $\mathbf{m}(\mathbf{r}) = \mathbf{M}(\mathbf{r})/M_s$ on the *z*-coordinate along the dot thickness, and consider instead a two dimensional magnetization distribution. The vortices bear topological charges referred to vorticity. The vorticity is the degree of mapping[6] $\mathbf{m}(x,y)$ of the *xOy* plane to the surface of a unit sphere $\mathbf{m}^2 = 1$. In general, the vortex can be described by three integer indices: the polarization (*p*), chirality (*C*) and vorticity (*q*). The vortex core polarization (*p*=±1) is defined as direction of the $m_z$ component in the vortex center (vortex core). The chirality is the direction of rotation of the vortex magnetization with respect to the dot center. The vortex low-frequency dynamic susceptibility and effective damping are calculated below assuming the topologically simplest vortex (*q*=1). We explicitly account for the dot magnetostatic energy due to volume and surface magnetic charges. This is because it determines the response of the vortex on the external magnetic field for submicron sized dots within the 10 MHz - 1 GHz frequency range. The relation to the theory of ferromagnetic resonance is demonstrated.

To investigate vortex dynamics in thin magnetic particles we use the effective equation of motion for vortex collective coordinates[19] which can be derived from the Landau-Lifshitz equation of motion. We consider the vortex translation modes, where the vortex center undergoes oscillations around its equilibrium position. These modes have the lowest frequencies in the vortex excitation spectra for finite, thin dots. We use the angular parameterization $(\Theta, \Phi)$ for the dot magnetization components $m_z = \cos\Theta$, $m_x + im_y = \sin\Theta \exp(i\Phi)$. For the static case $\Theta = \Theta_0(\rho)$, and $\Phi = q\varphi + C\pi/2$, where $\rho, \varphi$ are the polar coordinates. The Lagrangian corresponding to the Landau-Lifshits equation of motion is:

$$\Lambda = L\int d^2\boldsymbol{\rho} \left[ \frac{M_s}{\gamma} \dot{\Phi}(1-\cos\Theta) - w(\Theta,\Phi) \right], \tag{1}$$

where the energy density $w = A[(\nabla\Theta)^2 + \sin^2\Theta(\nabla\Phi)^2] + w_m + w_H$, $w_m = -\mathbf{M}\cdot\mathbf{H}_m/2$ is the magnetostatic energy density, $\mathbf{H}_m$ is the magnetostatic field, $w_H = -\mathbf{M}\cdot\mathbf{H}$ is the Zeeman energy density, γ is the gyromagnetic ratio, and the dot denotes the derivative with respect to time. The first term is the exchange energy, where $A$ is the exchange stiffness.

The vortex center behaves as a particle and can be characterized by its coordinate **X**, mass, momentum *etc*. But the parameters of the equations of motion depend on the vortex magnetization distribution and vortex topological charges. In order to describe the translation modes of the vortex motion, we use a collective-variable approach in the form of the generalized Thiele's equation,[4] which accounts for the vortex mass. Assuming that the vortex magnetization moves in the form of traveling-wave $\mathbf{M}(\mathbf{x},t) = \mathbf{M}[\mathbf{x} - \mathbf{X}(t)]$ and keeping only terms linear and quadratic in the vortex velocity $\dot{\mathbf{X}}$, the Lagrangian (1) can be written in the convenient form:

$$\Lambda(\mathbf{X},\dot{\mathbf{X}}) = \frac{1}{2}\dot{\mathbf{X}}\hat{M}\dot{\mathbf{X}} + \frac{1}{2}(\mathbf{G}\times\mathbf{X})\cdot\dot{\mathbf{X}} - W(\mathbf{X}), \qquad (2)$$

and the corresponding equation of motion is

$$\hat{M}\ddot{\mathbf{X}} - \mathbf{G}\times\dot{\mathbf{X}} + \frac{\partial W(\mathbf{X})}{\partial \mathbf{X}} = 0, \qquad (3)$$

where $\mathbf{X} = (X, Y)$ is the vortex center position, $W(\mathbf{X})$ is the potential energy of the vortex shifted from its equilibrium position at $\mathbf{X}=0$, and $\hat{M}$ is the vortex mass, which, in general, is a tensor.

Equation (3) can be derived immediately from the Landau-Lifshitz equation. The first term in Eq. (3) is the analogue to the classical acceleration term. The second term is the gyroforce, determined by the vortex non-uniform magnetization distribution (topological charge).[19] The gyroforce is proportional to the gyrovector $\mathbf{G} = -G\hat{\mathbf{z}}$, where $G = 2\pi qpLM_s/\gamma$, and $\hat{\mathbf{z}}$ is the unit vector perpendicular to the dot plane *xOy*. The gyrovector and vortex mass can be calculated from their definitions.[19] The non-zero gyrovector is of principal importance for the vortex dynamics description. It can be calculated by integration over the vortex core[8] (where $\nabla\Theta \neq 0$).

The restoring force [third term in Eq. (3)] appears due to the finite dot size in-plane and is directed toward the dot center. For submicron dot radii, the dot magnetostatic energy (note that a shifted vortex induces magnetic charges) provides the main contribution to $W(\mathbf{X})$. To properly describe the vortex motion we need to introduce a reasonable description of the vortex shifted from the center ($\mathbf{X}=0$) of the particle. We use the "side charges free" model of the shifted vortex, which satisfies the magnetostatic boundary conditions $(\mathbf{m}\cdot\mathbf{n})_S = 0$ on the dot side surface S. The model is applicable to describe the vortex state stability with respect to a change of the dot geometrical parameters[5] ($R$ and $L$) and the vortex low-frequency dynamics.[8] It was shown in Ref. 20 that $(\mathbf{m}\cdot\mathbf{n})_S = 0$ is satisfied with good accuracy for thin dots where $L<<R$. The vortex energy dependence on $\mathbf{X}$ in cylindrical dots was calculated[8] on the basis of the model. The volume averaged magnetization of the shifted vortex is proportional to its displacement $\langle\mathbf{m}(\mathbf{r})\rangle_V = -\xi C\hat{\mathbf{z}}\times\mathbf{s}$, where $\xi = 2/3$ for this model, and $\mathbf{s}=\mathbf{X}/R$. For $s<<1$ one can write $W(\mathbf{X})=W(0)+\kappa X^2/2+O(X^4)$, where the stiffness coefficient $\kappa$ is a function of $R$ and $L$.[8] For a non-zero external field (which is variable, but must be spatially uniform) we get $W_H = \xi(CM_s/R)[\hat{\mathbf{z}}\times\mathbf{H}]\cdot\mathbf{X}$.

To calculate the dynamic susceptibility of the moving vortex we need to add to Eq. (3) a damping term having the form as suggested by Thiele[19] $\mathbf{F}_d = \hat{D}\dot{\mathbf{X}}$, where the damping dyadic is:

$$D_{\alpha\beta} = -\alpha_{LLG}\int d^2\boldsymbol{\rho}\left(\frac{\partial\Theta}{\partial x_\alpha}\frac{\partial\Theta}{\partial x_\beta} + \sin^2\Theta\frac{\partial\Phi}{\partial x_\alpha}\frac{\partial\Phi}{\partial x_\beta}\right), \quad \alpha,\beta = x,y, \qquad (4)$$

and $\alpha_{LLG}$ is the Gilbert damping parameter.

It can be shown that for cylindrical dots the damping tensor $D_{\alpha\beta} = D\delta_{\alpha\beta}$ is diagonal, and the damping constant $D$ is simply proportional to the total exchange energy of the dot. We do not assume microscopic damping mechanism, i.e. we consider $\alpha_{LLG}$ as a phenomenological parameter. Calculations within the soliton model of the magnetic vortex[4,21] yield $D = -\alpha_{LLG}\pi M_s L(2 + \ln(R/b))/\gamma$, where $b = b(L)$ is the thickness dependent vortex core radius. $b(L) = 0.68 L_e (L/L_e)^{1/3}$ at $L \geq L_e$ within the model[21], and $L_e = (2A/M_s^2)^{1/2}$ is the exchange length. The linearized equation of motion of the vortex center $\mathbf{X}(t)$, including the relaxation and Zeeman terms, is:

$$\hat{M}\ddot{\mathbf{X}} - \mathbf{G} \times \dot{\mathbf{X}} - \hat{D}\dot{\mathbf{X}} + \kappa \mathbf{X} + \mu(\hat{\mathbf{z}} \times \mathbf{H}) = 0, \quad \mu = \xi C \frac{M_s}{R}. \tag{5}$$

For the linear vortex motion all the coefficients in Eq. (5) should be calculated at $\mathbf{X} = 0$, and $\dot{\mathbf{X}} = 0$. The parameters $\mathbf{G}$ and $\hat{D}$ are properties of the static vortex, whereas $\hat{M}$, $\kappa$ and $\mu$ depend on the magnetization distribution in a moving (shifted from the equilibrium) vortex. (It can be shown that the mass term could be neglected for typical dot sizes.)

To find the response of the average dot magnetization on uniform variable external field $\mathbf{H}(t)$, i.e. dynamic susceptibility tensor, we use the relation $\overline{\mathbf{M}}(t) = \langle \mathbf{M}(\vec{r},t)\rangle_V = -\mu(\hat{\mathbf{z}} \times \mathbf{X}(t))$ and calculate the response of $\mathbf{X}(t)$ neglecting the high-frequency vortex modes. We consider two particular cases of the driving field $\mathbf{H}$: (i) circularly and (ii) linearly polarized fields oscillating with frequency $\omega$ in the dot plane $xOy$. Then the motion can be described as that of the in-plane magnetization components $\overline{M}_{x,y}(t)$ keeping the normal component $\overline{M}_z$ as a constant, i.e. similar to uniform precession in an isotropic infinite ferromagnet in magnetic field along the $Oz$ axis. Solving Eq. (5) and using the definition $\overline{\mathbf{M}}(\omega) = \chi(\omega)\mathbf{H}(\omega)$, where $\mathbf{H}(t) = \mathbf{H}(\omega)\exp(i\omega t)$, we get for the circular dynamic susceptibility the expression:

$$\chi(\omega) = \chi(0)\frac{\omega_0(\omega_0 - \omega + id\omega)}{(\omega_0 - \omega)^2 + d^2\omega^2}, \quad \omega_0 = \frac{\kappa}{G}, \quad d = -\frac{D}{G}. \tag{6}$$

Here $\chi(0)$ is the static susceptibility of the model depending on $R$ and $L$.[8] The frequency dependence of this susceptibility coincides with one for the case of ferromagnetic resonance in an infinite ferromagnet,[22] except that the effective damping parameter is $d = \alpha_{LLG}(1+\ln(R/b)/2)$ and the resonance frequency $\omega_0$ has a different physical sense and value. The size dependence of the eigenfrequency $\omega_0$ was calculated in Ref. 8. The logarithmic term is not small for typical dot sizes $R$=0.3-1 μm, $L$=10-100 nm, and $d \approx 3\alpha_{LLG}$. Note that if $\omega > 0$ (the right polarization of the rotating a.c. field) the susceptibility (6) has resonance character only for $p > 0$, when $\omega_0 = |\omega_0|p$ is positive. I.e., circularly polarized field excites the vortex mode only if $p\omega > 0$. The calculated value of the magnetization component $\overline{M}_z = 0.39(b/R)^2 M_s$ is negligibly small due to small core radius $b \sim L_e$=20 nm and decreases as the vortex trajectory radius $|\mathbf{X}(t)|$ increases.[14] The reason is that there is a distortion of the moving vortex profile, until the reversal of its core at some critical a.c. field magnitude, as was demonstrated experimentally by Puzic et al.[12] For the moving vortex $\overline{M}_z$ is much smaller than the $M_z$ component for the ferromagnetic resonance case, where $M_z \cong M_s$ in the linear regime. For a linearly polarized field $\mathbf{H}(t)$, let us say along $Ox$ axis, we can calculate the $xx$-component of the dynamic susceptibility. Experimentally the measured lineshape of the vortex resonance for a circular magnetic dot is proportional to the imaginary part of the in-plane susceptibility:

$$\operatorname{Im}\chi_{xx}(\omega) = d\chi(0)\frac{\omega\omega_0\left[\omega_0^2 + (1+d^2)\omega^2\right]}{\left[\omega_0^2 - (1+d^2)\omega^2\right]^2 + 4d^2\omega_0^2\omega^2}. \qquad (7)$$

The effective damping parameter $d$ is typically small and the dependence of the susceptibility (7) on frequency reveals a sharp Lorentz-like peak at $\omega \approx \omega_0$ with a maximum value of $\operatorname{Im}\chi_{xx}(\omega_0) \cong \chi(0)/2d$. The resonance line half-width is $\Delta\omega = 2d\omega_0$. The vortex resonance was experimentally observed in Ref. 14 for different values of *R* and *L* in permalloy circular dots. The value of the bare damping parameter $\alpha_{LLG}$ can be extracted from the linewidth assuming that the measured signal is proportional to $\text{Im}\,\chi_{xx}(\omega)$.[14] The resonance frequencies and linewidths were measured for dot arrays with (*R*, *L*)=(550 nm, 40 nm), (500 nm, 100 nm),[23] (1100 nm, 40 nm), (1000 nm, 20 nm), correspondingly: $(\omega_0, \Delta\omega)$=(271 MHz, 15 MHz), (521 MHz, 30 MHz), (161.2 MHz, 10.9 MHz), (83.2 MHz, 2.2 MHz).[14] The line shape Eq. (8) is compared to experimental data[14] in Fig. 2. It is convenient to operate with the ratio $\Delta\omega/\omega_0$, which does not depend on the particular model of the moving vortex. Using the experimentally measured ratio $\Delta\omega/\omega_0$, the value of the damping parameter $\alpha_{LLG}$ can be determined for each dot array. We get the values $\alpha_{LLG}$=9.9 $10^{-3}$, 11.3 $10^{-3}$, 10.8 $10^{-3}$, 4.2 $10^{-3}$ for the corresponding dot arrays. The first three values almost coincide and are close to the value of $\alpha_{LLG} = 0.008$ for bulk permalloy.[22] The fourth value is approximately two and half times less and will be discussed elsewhere.

We have derived expressions for the dynamic susceptibility of the vortex state in soft magnetic cylindrical dots as a function of size. The dynamical susceptibility has a form that is similar to the one for ferromagnetic resonance in an infinite ferromagnet. But the parameters of the dynamic susceptibility ($\chi(0)$, $\omega_0$ and *d*), that determine the resonance line position and shape depend on the geometrical size of the dots. The damping parameter $\alpha_{LLG}$ can be found quantitatively from the vortex resonance experimental data on the basis of the derived equations. The calculated values of $\alpha_{LLG}$ for micron-diameter FeNi dots are 0.010-0.011. The calculations presented can be readily extended for other dot shapes, such as elliptical, square, *etc*.

The author thanks K. Buchanan and V. Novosad for supplying experimental data on the vortex resonance in FeNi circular dots, and S.D. Bader for stimulating discussions. This work was supported by the U.S. Department of Energy, BES Material Sciences under Contract No. W-31-109-ENG-38.

**Figure captions**

Fig. 1. Geometry of the problem: circular cylindrical dot of radius *R* and thickness *L* with the magnetic vortex ground state.

Fig. 2. Vortex low-frequency resonance intensity vs. frequency in a circular FeNi dot (*R*=550 nm, *L*=40 nm): the black line is the experimental data[14]; the red line is the normalized susceptibility $\mathrm{Im}\,\chi_{xx}(\omega)2d/\chi(0)$ calculated by Eq. (7) with the damping parameter $\alpha_{LLG}$=9.9 10$^{-3}$. .

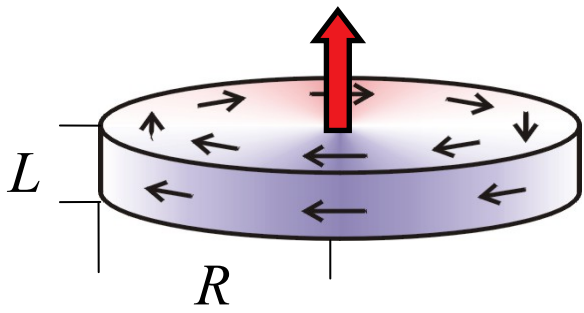

Fig. 1 to the manuscript by K.Y. Guslienko "Low-frequency vortex…"

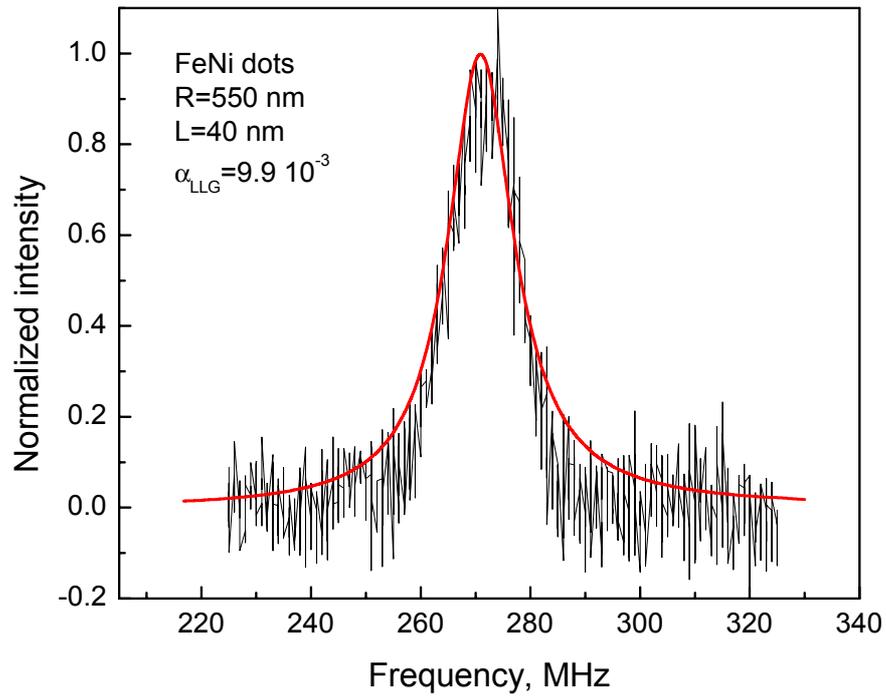

Fig. 2 to the manuscript by K.Y. Guslienko "Low-frequency vortex…"